\documentstyle[preprint,aps,floats,amssymb,epsfig]{revtex}

\newcommand{\be}{\begin{equation}}
\newcommand{\ee}{\end{equation}}
\newcommand{\bea}{\begin{eqnarray}}
\newcommand{\eea}{\end{eqnarray}}
\newcommand{\bml}{\begin{mathletters}}
\newcommand{\eml}{\end{mathletters}}

\begin{document}
\tighten
\preprint{DCPT-03/43}
\draft




\title{Spinning Solitons of a Modified Non-Linear Schr\"odinger equation}
\renewcommand{\thefootnote}{\fnsymbol{footnote}}
\author{Yves Brihaye\footnote{Yves.Brihaye@umh.ac.be}}
\address{Facult\'e des Sciences, Universit\'e de Mons-Hainaut,
7000 Mons, Belgium}
\author{Betti Hartmann\footnote{Betti.Hartmann@durham.ac.uk}}
\address{Department of Mathematical Sciences, University
of Durham, Durham DH1 3LE, U.K.}
\author{Wojtek J. Zakrzewski\footnote{W.J.Zakrzewski@durham.ac.uk}}
\address{Department of Mathematical Sciences, University
of Durham, Durham DH1 3LE, U.K.}
\date{\today}
\setlength{\footnotesep}{0.5\footnotesep}

\maketitle
\begin{abstract}
We study soliton solutions of a modified non-linear Schr\"odinger (MNLS) equation.
Using an Ansatz for the time and azimuthal angle dependence previously considered
in the studies of the spinning Q-balls, 
we construct multi-node solutions of MNLS as well as spinning generalisations.
\end{abstract}

\renewcommand{\thefootnote}{\arabic{footnote}}
\section{Introduction}
Solitons and instantons \cite{raja} are classical, non-singular, finite energy solutions
of non-linear field equations. There exist topological solitons and non-topological
solitons. The former are characterised by a conserved topological charge which
results (in most cases) from a spontaneous symmetry breaking of the theory.
Examples are monopoles, vortices and domain walls. Non-topological solitons \cite{lee}
have a conserved Noether charge which results -in contrast to
the case of topological solitons - from a symmetry of the Lagrangian.
Examples of these are so-called Q-balls \cite{qballs} which are solutions of
a complex scalar field theory involving a $\phi^6$ potential.

The topic of classical field equations' solutions 
that are rotating has gained a lot of interest
in recent years. It is well known that the inclusion of gravity can lead to
a number of rotating solutions such as the famous Kerr-Newman family
of rotating black holes as well as globally regular gravitating solutions
namely so-called boson stars \cite{bs}. These are spinning solutions
of the coupled Einstein and Klein-Gordon equations. We refer to them
as spinning rather than rotating since they are not rotating in the sense
of classical mechanics, but have a time and azimuthal angle dependence
of the form $\exp(i\omega t+iN\theta)$. 

However, classical solutions in flat space that are spinning
have only been constructed very recently \cite{volkov}. These are Q-balls
which were rotated in the sense that they have the time 
and azimuthal angle dependence chosen for boson stars.
Solutions with $k=0,1,2$ nodes of the scalar field and $N=1,2,3$ have been
constructed nuemrically. These solutions are of particular interest because
since, as it has been shown in \cite{volkov2}, the topological solitons in SU(2) gauge theory
cannot rotate. 

In this paper, we study spinning solitons of a modified non-linear Schr\"odinger
equation (MNLS) which arises, in the ontinuum limit, in the studies
of solitons on discrete quadratic \cite{bpz1,bpz2,bpz3}
and hexagonal lattices \cite{hz}, respectively. 
These equations were studied recently for $d\ge 2$ space dimensions in a recent paper \cite{bepz}.
In the present paper, we reconsider in detail the case $d=2$ adopting the viewpoint
of \cite{volkov}. Namely, we have managed to construct excited solutions for both
non-spinning and spinning solitons. Physical quantities characterising the excited
solutions are compared to those characterising the fundamental solutions.

Our paper is organised as
follows: in Section II we define the model, in section III we study 
its non-spinning solutions, while in section IV we concentrate our attention on the spinning
generalisations. We present our conclusions in section V.

\section{Modified non-linear Schr\"odinger equation}
The modified nonlinear Schr\"odinger (MNLS) equation in 2 dimensions was presented in  
\cite{bpz1,bpz3,bpz2,hz} and reads:
\begin{equation}
\label{cnls}
i\frac{\partial \psi}{\partial t}+\Delta \psi +
ag\psi\left(\vert\psi\vert^2+ 
b\Delta \vert\psi\vert^2\right)=0  \ .
\end{equation}
The constants $a$ and $b$ depend on the underlying
geometry of the lattice, where $a=2$, $b=1/12$ for the quadratic lattice
\cite{bpz1,bpz3,bpz2} and $a=4$, $b=1/8$ for the hexagonal lattice 
\cite{hz}. In the following, we will choose $a$, $b$ according to
the hexagonal lattice.

$g$ is the coupling constant of
the system
and $\Delta$ denotes the Laplacian in $2$ dimensions. 
The equation (\ref{cnls}) can be derived from the functional defined by the following Lagrangian density:
\begin{equation}
\label{lag}
{\cal L}=i\left(\frac{\partial \psi^*}{\partial t}\psi-
\frac{\partial \psi}{\partial t}\psi^*\right)+|\vec{\nabla}\psi|^2-
2g |\psi|^4+\frac{1}{4}g (\vec{\nabla} |\psi|^2)^2 \ .
\end{equation}
(In fact $-\cal L$ is the Lagrangian density but we use the convention above throughout the paper.)
The system has many conserved quantities \cite{bepz}, namely the space integral of the static part of the
Lagrangian (\ref{lag}) which can be interpreted as the ``static energy'':
\begin{equation}
E=\int \left( |\vec{\nabla}\psi|^2-2g |\psi|^4 + 
\frac{1}{4}g(\vec{\nabla} |\psi|^2)^2\right)dxdy \ 
\end{equation}
as well as the norm $n$ of the solution:
\begin{equation}
n^2=\int \psi \psi^* dx dy  \ .
\end{equation}
We notice that by rescaling $\psi\rightarrow\psi/\sqrt{g}$, we can scale out $g$ from
(\ref{cnls}). (Of course, $E$ and $n$ are also rescaled then.) 

\section{Non-spinning solitons}
\subsection{Ansatz}

$\psi$ is a complex scalar field, we choose first to consider it to be given by
the (radially symmetric) Ansatz:
\begin{equation}
\label{ansatz1}
\psi(t,r)=e^{i\omega t} \phi(r) \ 
\end{equation}
where $r=\sqrt{x^2+y^2}$.
First, we remark that the quantity
\begin{equation}
Q=\frac{1}{i}\int \left(\psi^*\frac{\partial \psi}{\partial t}
-\psi\frac{\partial \psi^*}{\partial t}\right)dxdy
\end{equation}
is conserved for the Ansatz (\ref{ansatz1}), since
\begin{equation}
\label{eq}
Q=2\omega n^2 \ .
\end{equation}
So, the action of the solutions constructed within the above Ansatz appears as a sum of two conserved quantities.

Inserting the Ansatz (\ref{ansatz1})  into (\ref{cnls}), we obtain the following ordinary differential equation:
\begin{equation}
\label{eq1}
\phi''+\frac{\phi'}{r} + \frac{g\phi\phi'^2}{1+g\phi^2}+\frac{4g\phi^3-
\omega\phi}{1+g\phi^2}=0 \ ,
\end{equation}
where the prime denotes the derivative with respect to $r$.
This equation can also be derived from the following effective one-dimensional Lagrangian density:
\begin{equation}
\label{sL}
{\cal L}_{ns}=(1+g\phi^2)\phi'^2-2g\phi^4+\omega\phi^2 \ .
\end{equation}
The form of the potential $V(\phi)=2g\phi^4-\omega\phi^2$ is given 
for $g=10$ and $\omega=1$, $0$, $-1$ in Fig.~1. If we choose either $g \leq 0$ or $\omega \leq 0$, this potential
has no local maxima. In a classical mechanics analog, we can think of this as a 
particle with position dependent mass in a $x^4$ potential.
 In order to have localised, finite energy solutions the function $\phi(r)$ has to approach
 a zero of the potential for $r\rightarrow\infty$. This means here that $\phi(r\rightarrow \infty)\rightarrow 0$.
If we come back to the analog with the particle, this means that (with $r$ being now the ``time'')
the particle starts of with zero velocity $\phi'(0)=0$ (this in fact is one of the used boundary conditions).
It should end up at the local maximum of the potential (see Fig.~1). Employing the same argument
as in Ref.\cite{volkov}. the form of the potential suggests further that solutions with $k=1,2,...$ nodes in the function
$\phi(r)$ should also exist for $\omega >0$.

It is convenient to also define the total energy of the solution according to:
\begin{equation}
E_{tot}=\int {\cal L}_{ns} r dr d\theta \ .
\end{equation}
Our numerical results show that $E_{tot}$ is always positive (see the Numerical
results  section below) and $E_{tot}(k=0) \le E_{tot}(k >0)$.

Before we discuss the numerical results we present the analytical behaviour of the solutions
around the origin and at infinity.
Close to the origin $r=0$, the function $\phi(r)$ behaves like:
\begin{equation}
\phi(r << 1)=C_0 + O(r^2) \ \ , \ \ C \ \ {\rm constant} \ .
\end{equation}
The asymptotic behaviour of the function $\phi(r)$ can easily be
determined from (\ref{eq1}). The linearized equation is indeed a Bessel
equation and  leads to the following asymptotic behaviour: 
\begin{equation}
\label{besselj0}
\phi(r>>1) = J_0(x) 
\sim \frac{\cos(x-\pi/4)}{\sqrt x} \ \ , \ \ {\rm if} \ \omega < 0
\end{equation} 
\begin{equation}
\label{besselk0}
  \phi(r>>1) = K_0(x)
 \sim \frac{\exp(-x)}{\sqrt{x}} \ \ , \ \ {\rm if} \ \omega > 0
 \end{equation}
where $x \equiv \sqrt{\vert \omega \vert} r$.  
The function $\phi(r)$ therefore oscillates around $\phi = 0$
for $\omega < 0$ and decays exponentially  for $\omega > 0$.
Both type of behaviours are confirmed by our numerical analysis.

Note that if the factor $(1+g\phi^2)$ is replaced by $1$, 
the Lagrangian (\ref{sL}) reduces to that
of a scalar field in a $\phi^4$ potential. This is known to have
no stable solutions, thus a $\phi^6$ potential was introduced \cite{qballs} in this case.
In our model no such $\phi^6$ term is possible, but we have a non-constant
prefactor of the kinetic term, which leads to the existence of
stable solutions. In the non-linear $\sigma$-model also a term
in front of the kinetic term appears which results from
the fact that the scalar field is constrainted. 

To construct soliton solutions of our model, we introduce the following boundary conditions
for non-spinning solutions \cite{hz}:
\begin{equation}
\partial_r\phi(r=0)=0 \ \ , \ \ \phi(\infty)=0 \ .
\end{equation}
With these boundary conditions, the function $\phi(r)$ can cross the $r$-axis
$k$ times, i.e. multi-node solutions are possible.

\subsection{Numerical results}

Without loosing generality, we can set $g=1$ in the following.

In Fig.~2, we present the two conserved quantities $E$, $n$ and the value of $\phi$ at the origin, $\phi(0)$,
as function of $\omega$ for the non-spinning, zero-node soliton solution. 
We find that the norm for $\omega << 1$ tends to a finite value as $\omega\rightarrow 0$.
It is interesting to connect this result with the one of \cite{hz}.
With respect to this paper, we have rescaled $\psi\rightarrow \psi/\sqrt{g}$, so that
the norms are related by $n=\sqrt{g}\tilde{n}$, where $\tilde{n}$ denotes the norm of the unscaled
solution. In \cite{hz}, the convenion $\tilde{n}=1$ was adopted, thus $n=\sqrt{g}$
and in the limit $\omega\rightarrow 0$, the critical values of the norm is equal to
$\sqrt{g_{cr}(N=0,k=0)}\approx \sqrt{2.93} \approx 1.71$, which is exactly what we find.

In \cite{volkov}, solutions were shown to exist only for a specific range of $\omega$.
Here, we find no upper or lower bound on $\omega$ as long as $\omega\ge 0$.
For $\omega < 0$, the equation becomes of a Bessel type and the solutions become oscillatory.
The difference between \cite{volkov} and our results is due to the fact
that in \cite{volkov} a $\phi^6$-potential was introduced,
while we have a non-constant prefactor in front of the kinetic term.
The $\phi^6$-potential leads to the restriction in $\omega$, while
in our model the $\phi^4$-potential is not subject to such restrictions. 

The first excited solution has a maximum at $r=0$. It
then crosses zero at $r=r_0$ and attains a minimum at , say $r = r_m$.
The numerical results show that   $\phi(r_m) \sim -\phi(0)/2$
and $r_m \sim  2 r_0  $.
For $\omega >> 1$ both $r_0$ and $r_m$ decrease, so that the
soliton becomes increasingly concentrated around its center. 
If $\omega << 1$ these
values get larger and the soliton is more delocalised. 
Since the norm of the first excited solution also tends to a finite value
in the limit $\omega\rightarrow 0$, we can conclude (following the above arguments
for the fundamental solution and adopting the conventions of \cite{hz}) that
the first excited solution exists only for a sufficiently high value of $g$, namely
for $\sqrt{g} > \sqrt{g_{cr}(N=0,k=1)}\approx 4.38$. This has to be compared to
$\sqrt{g_{cr}(N=0,k=0)}\approx 1.71$ of the fundamental solution.

Finally, we present the total energy $E_{tot}$ of the solutions
in Fig.~5. Clearly, the energy stays positive for all $\omega \ge 0$
and the zero-node ($k=0$) solution  has for all values
of $\omega$ smaller energy than that of the first excited ($k=1$) solution.

\section{Spinning solitons}
To construct spinning generalisations, we take the following Ansatz \cite{volkov}:
\begin{equation}
\psi=\phi(r)e^{i\omega t+iN\theta} \ , \ \ N \ \ {\rm integer} \ .
\end{equation}
The equation then takes the form:
\begin{equation}
\phi''+\frac{\phi'}{r} + \frac{g\phi\phi'^2}{1+g\phi^2}+\frac
{4g\phi^3-\omega\phi}{1+g\phi^2}
-\frac{N^2 \phi}{r^2(1+g\phi^2)}=0 \ .
\end{equation}

As before, this equation can equally be derived from the following static Lagrangian:
\begin{equation}
{\cal L}_{s}=\frac{1+g\phi^2}{2}\phi'^2-
g\phi^4+\frac{\omega}{2}\phi^2+\frac{N^2 \phi^2}{2r^2} \ .
\end{equation}
where now the potential becomes explicitely $r$-dependent due to the additional
term $-N^2 \phi^2/(2r^2)$.
Note that the equality (\ref{eq}) still holds for this Ansatz. The total energy $E_{tot}$ of the solution
again is the space-integral of the static Lagrangian ${\cal L}_{s}$.

For $r << 1$, the function $\phi(r)$ behaves as:
\begin{equation}
\phi(r << 1)= r^N\left(C_N+O(r^2)\right) \ \ , \ \ C_N \ \ {\rm constant} \ .
\end{equation}
Moreover, the asymptotic behaviour of $\phi(r)$ is now determined
by the function $J_N(x)$ and $K_N(x)$ respectively for $\omega < 0$
and $\omega > 0$.

As a consequence (\ref{cnls}) has to be solved subject to the following boundary conditions:
\begin{equation}
\phi(r=0)=0 \ \ , \ \ \phi(\infty)=0 \ .
\end{equation}
Again, the choice of these conditions allows for multinode solutions.

\subsection{Numerical results}

Like for the non-spinning solutions, we choose $g=1$.

The conserved quantities $E$, $n$ and the value of the derivative of
$\phi(r)$ at the origin, $\phi'(0)$, for the $N=1$ solutions
with no nodes ($k=0$) and one node ($k=1$), respectively, are shown in Fig.~3.
Again, the norm of the solutions tends to a finite value for $\omega\rightarrow 0$.
Thus, again employing the language of \cite{hz}, we conclude that the normalised spinning
solutions exist only for coupling constants $g$ larger than a critical value $g_{cr}(N,k)$.
Like in the non-spinning case, we find that the critical value of the $k=0$  solution
is smaller than that of the $k=1$ solution. In summary, we find for the critical values:
\begin{equation}
g_{cr}(0,0) < g_{cr}(1,0) < g_{cr}(0,1) < g_{cr}(1,1) \ .
\end{equation}

The profiles of the solutions with $N=1$ and $k=0$, $1$ 
are shown in Fig.~4 together with the two non-spinning solutions.
The functions $\phi(r)$ have a maximum at $r= r_M$ and we find that
the value of this maximum 
increases slowly as a function of $\omega$. Again the solution
and its first excitation gets narrower (resp. more spread out)
for $\omega >> 1$ (resp. $\omega << 1$).

In Fig.~5, we present the total energy $E_{tot}$ of the solutions
as a function of $\omega$. Again, like in the $N=0$ case, the energy is always
positive and the energy of the fundamental solution with no nodes
is smaller than that of the first excited solution.
Comparing the energy of the solution for fixed $k$ and different $N$,
we find that the non-spinning solution has lower energy than the
spinning solution, which is an expected result since the spinning adds energy to the
solution.

\section{Conclusions}
The study of  rotating solitons in flat space
is still a little studied problem in classical field theory.
Recently, non-topological spinning Q-balls have been constructed \cite{volkov}
using an Ansatz used to construct spinning boson stars \cite{bs}.

In this paper, we have considered  an equation which is related
to the continuum limit of a system 
of equations describing the interaction of a complex Schr\"odinger-like field
with a quadratic \cite{bpz1,bpz3,bpz2}, resp. hexagonal lattice \cite{hz}.
Although the system is originally  not
relativistically invariant, the radially symmetric
equations resemble classical
static equations of a non-linear sigma model type supplemented by
a ``symmetry breaking'' quartic potential.

The pattern of solutions of this non-linear equation seems extremely rich,
containing several families  of solutions (for this see also \cite{bepz}).
The solutions are mainly characterized by the parameter $\omega$
defining the harmonic time dependence. Remarkably, the function
determining the soliton tends to zero in the limit $\omega \rightarrow
0$  although the ``norm'' of the solution stays finite in the same limit.
This phenomenon provides a natural explanation of the fact that 
the solitons studied in  \cite{hz} exist only for large enough
values of the coupling constant.

Among possible generalisation of our
result we mention  the construction of non-spinning and spinning solutions (and their excited solutions)
of
a three dimensional version of the equation and/or the coupling
to a electromagnetic field, thus leading to charged solutions.\\
\\
\\
{\bf Acknowledgments} Y. B. gratefully acknowledges the Belgian F.N.R.S. for financial
support. B. H. was supported by an EPSRC grant.

\newpage
\begin{figure}
\centering
\epsfysize=10cm
\mbox{\epsffile{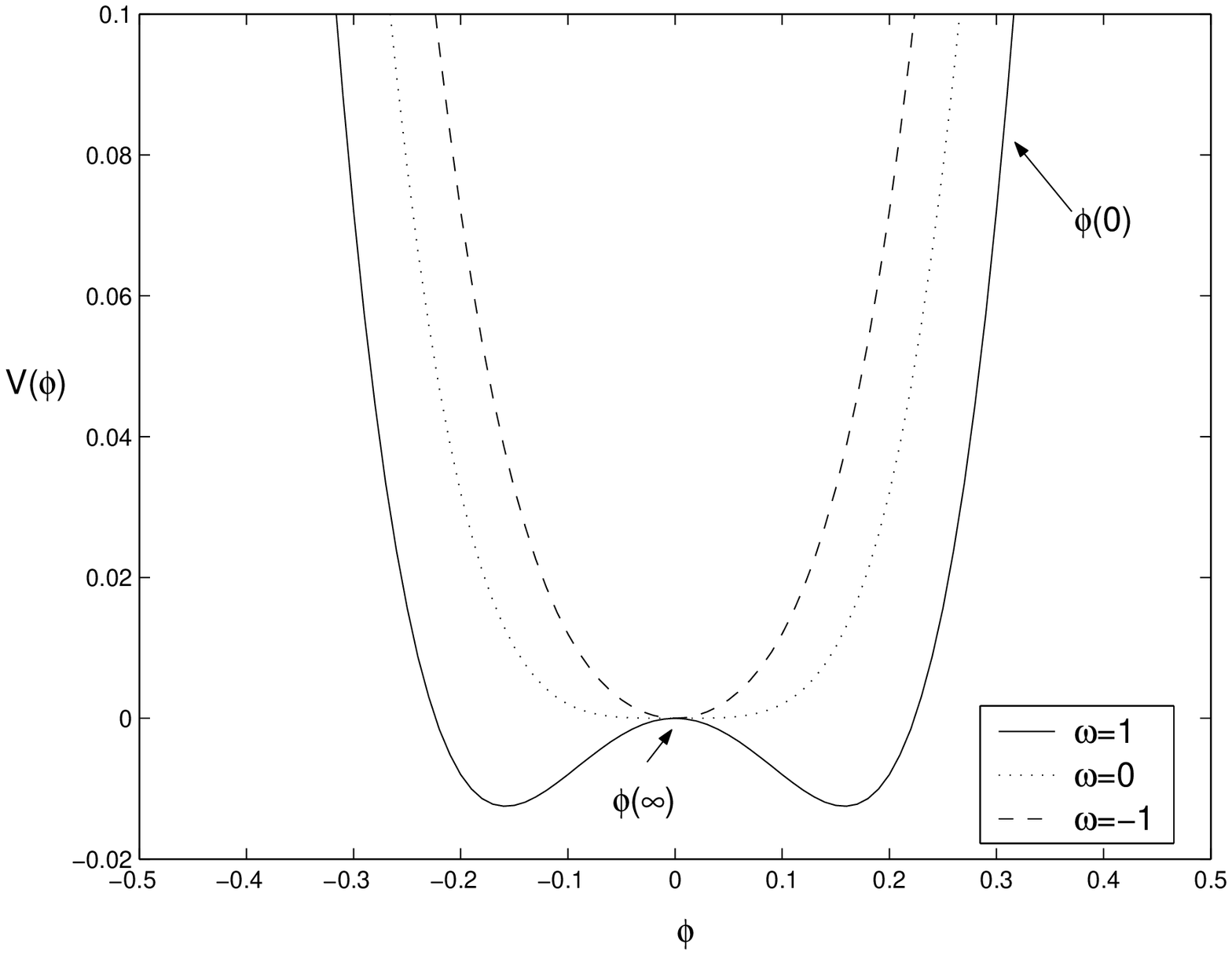}}
\caption{\label{Fig.1} The potential $V(\phi)$ is shown for $g=10$ and $\omega=1$.}
\end{figure}

\newpage
\begin{figure}
\centering
\epsfysize=10cm
\mbox{\epsffile{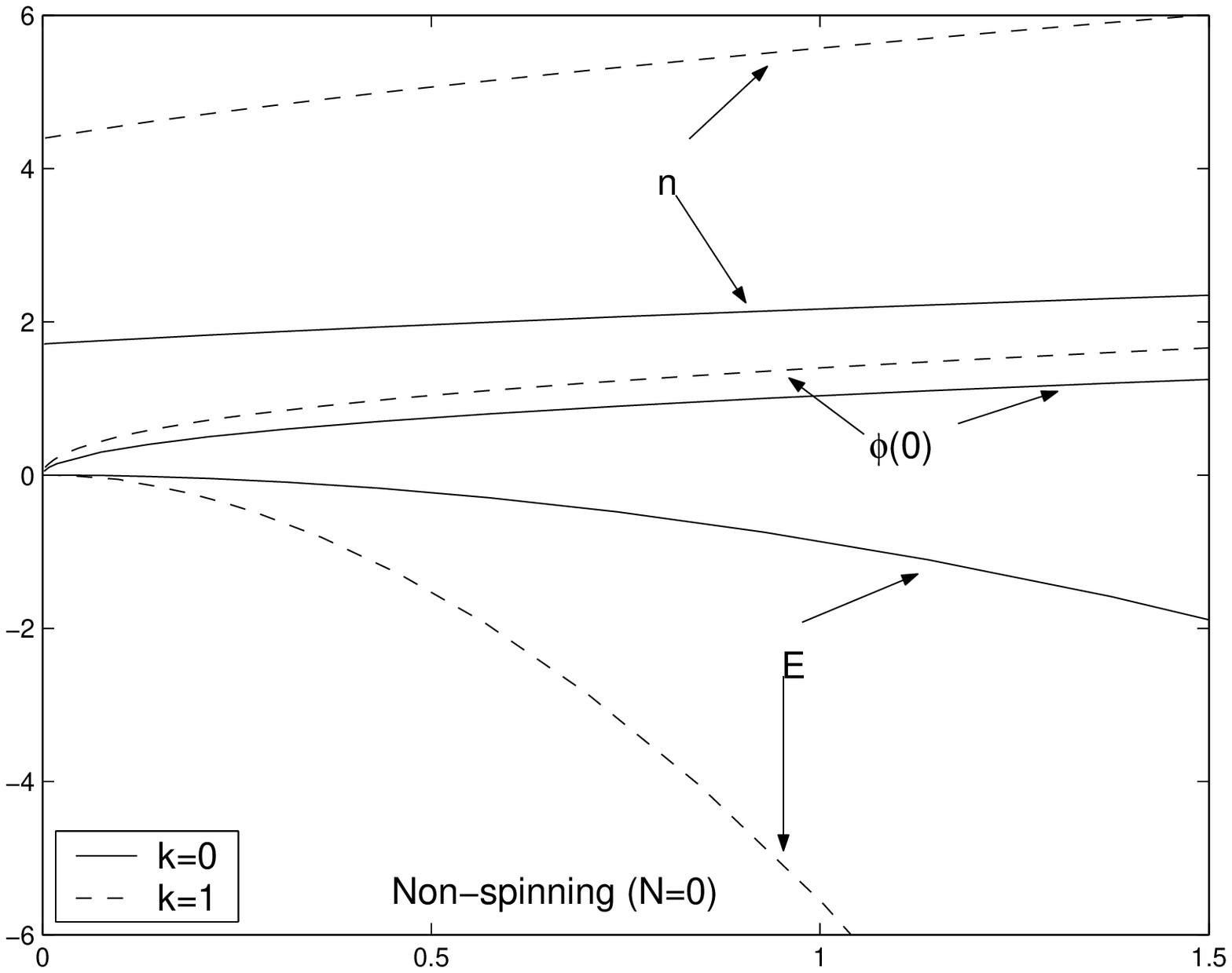}}
\caption{\label{Fig.2} The conserved quantities $E$, $n$ and the value $\phi(0)$ are shown as functions
of $\omega$ for the non-spinning solutions with $k=0$, $k=1$, respectively.}
\end{figure}

\newpage
\begin{figure}
\centering
\epsfysize=10cm
\mbox{\epsffile{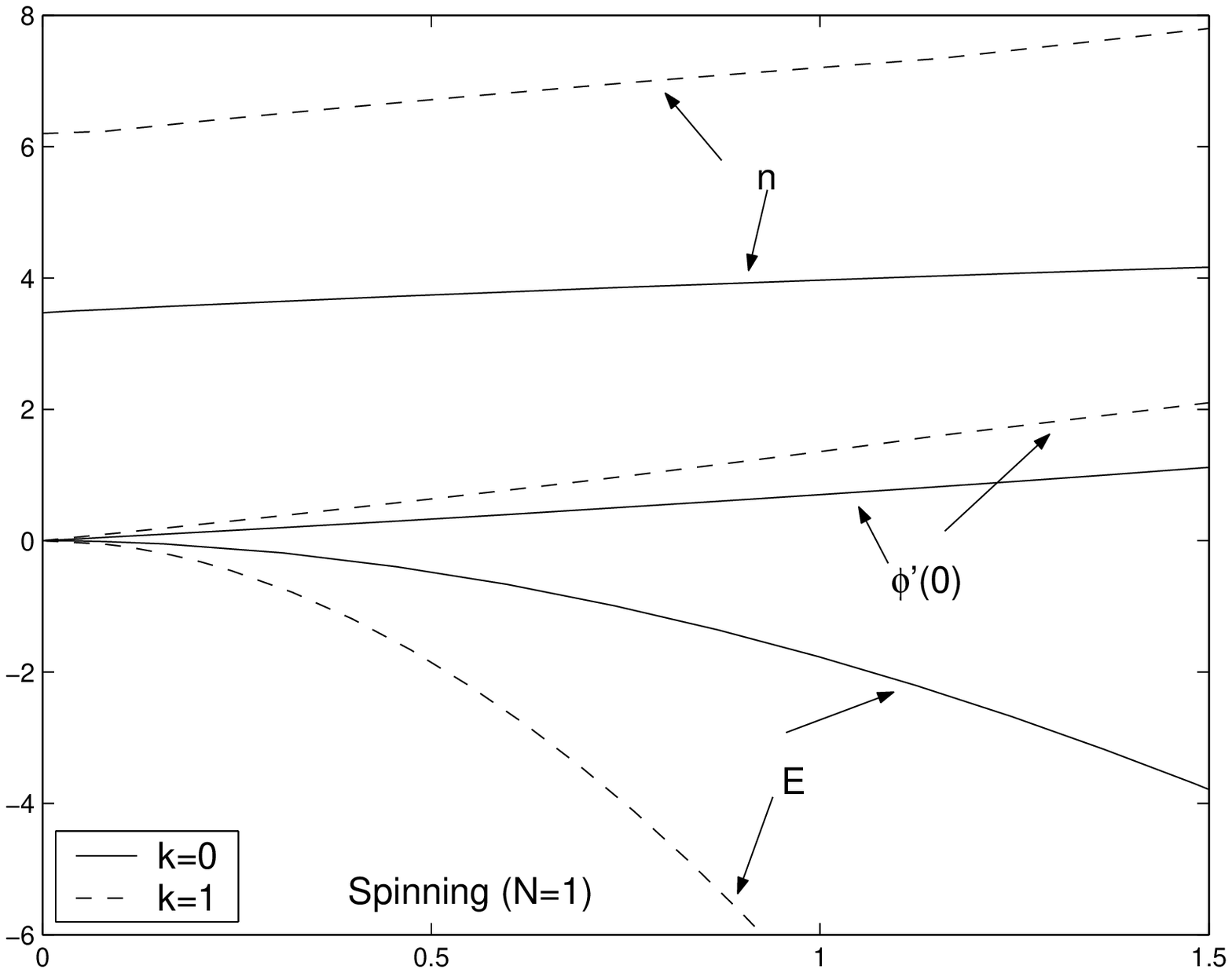}}
\caption{\label{Fig.3} The conserved quantities $E$, $n$  and the value $\phi'(0)$ are shown as functions
of $\omega$ for the spinning solutions ($N=1$) with $k=0$, $k=1$, respectively. }
\end{figure} 

\newpage
\begin{figure}
\centering
\epsfysize=10cm
\mbox{\epsffile{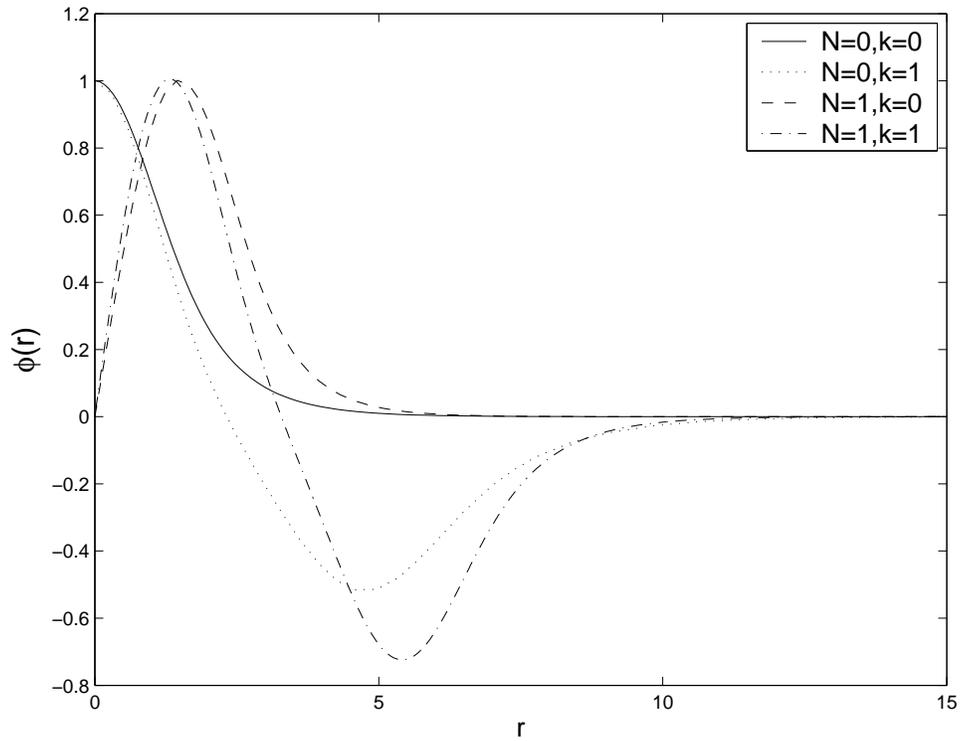}}
\caption{\label{Fig.4} The profiles $\phi(r)$ of the soliton solutions are shown for different choices of
the node number $k$ and the parameter $N$. $N=0$ are non-spinning solitons. $g=1$. }
\end{figure}

\newpage
\begin{figure}
\centering
\epsfysize=10cm
\mbox{\epsffile{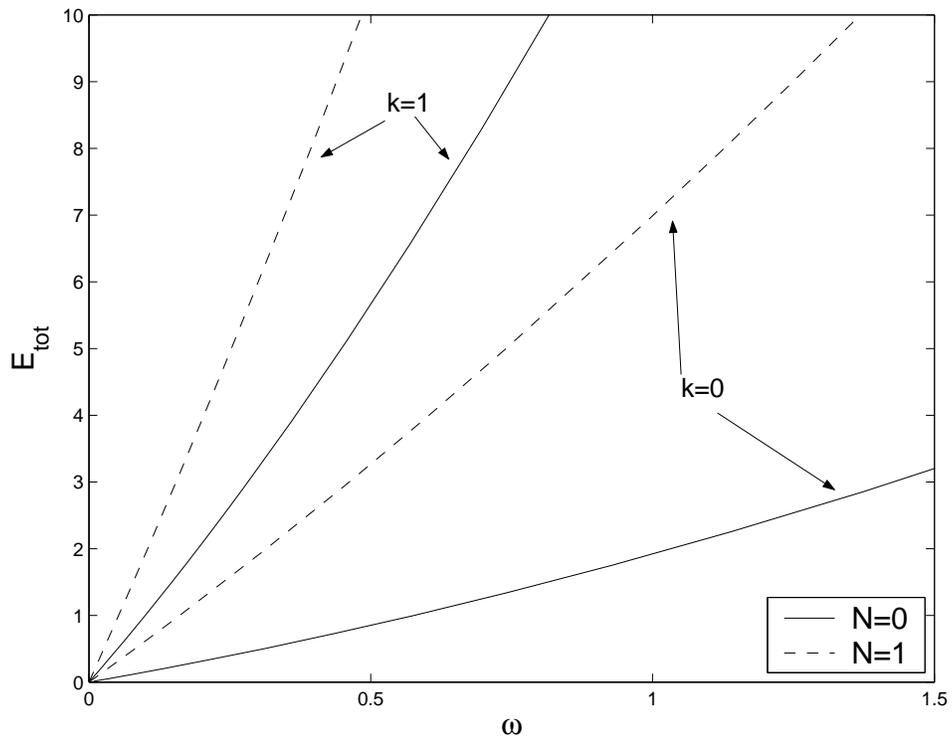}}
\caption{\label{Fig.5} 
The total energy $E_{tot}$ of the non-spinning ($N=0$) and the spinning
$(N=1)$ solutions with $k=0,1$ nodes is shown as function of $\omega$}
\end{figure}

\end{document}